\documentclass[conference]{IEEEtran}

\usepackage{graphicx}
\usepackage{algorithm}
\usepackage{algpseudocode}
\usepackage{cite}
\usepackage{verbatim}
\usepackage{amsmath}
\usepackage{url}
\ifCLASSOPTIONcompsoc
\usepackage[tight,normalsize,sf,SF]{subfigure}
\else
\usepackage[tight,footnotesize]{subfigure}
\fi

\begin{document}
%
\title{FlowCover: Low-cost Flow Monitoring Scheme in Software Defined Networks}

\author{\IEEEauthorblockN{Zhiyang Su, Ting Wang, Yu Xia, Mounir Hamdi}
\IEEEauthorblockA{Hong Kong University of Science and Technology \\
\{zsuab, twangah, rainsia, hamdi\}@cse.ust.hk}
}

\maketitle

\begin{abstract}
  Network monitoring and measurement are crucial in network management to facilitate quality of service routing and performance evaluation. Software Defined Networking (SDN) makes network management easier by separating the control plane and data plane. Network monitoring in SDN is light-weight as operators only need to install a monitoring module into the controller.
Active monitoring techniques usually introduce too many overheads into the network. The state-of-the-art approaches utilize sampling method, aggregation flow statistics and passive measurement techniques to reduce overheads. However, little work in literature has focus on reducing the communication cost of network monitoring. Moreover, most of the existing approaches select the polling switch nodes by sub-optimal local heuristics.
Inspired by the visibility and central control of SDN, we propose FlowCover, a low-cost high-accuracy monitoring scheme to support various network management tasks. We leverage the global view of the network topology and active flows to minimize the communication cost by formulating the problem as a weighted set cover, which is proved to be NP-hard. Heuristics are presented to obtain the polling scheme efficiently and handle flow changes practically. We build a simulator to evaluate the performance of FlowCover. Extensive experiment results show that FlowCover reduces roughly $50\%$ communication cost without loss of accuracy in most cases.
\end{abstract}

\IEEEpeerreviewmaketitle
\bstctlcite{BSTcontrol}

\section{Introduction}
Monitoring resource utilization is a common task in network management. Recently, as the rapid development of software defined networking (SDN), network management becomes easier and easier. A typical SDN based network consists of many switches and a logically centralized controller which monitors the whole network state and chooses routing paths. The separation of the control plane and data plane makes it possible to track the state of each flow in the control plane. Low-cost, timely and accurate flow statistics collection is crucial for different management tasks such as traffic engineering, accounting and intelligent routing.

There are two ways to measure the network performance: active or passive techniques. Active measurement obtains the network state by injecting probe packets into the network. Active measurement is flexible since you can measure what you want. It estimates the network performance by tracking how the probe packets are treated in the network. The accuracy is closely related to the probe frequency in general. However, the measurement packets will disturb the network, especially when sending measurement traffic with high frequency. In contrast to active measurement, passive measurement provides detailed information about the nodes being measured. For example, Simple Network Monitoring Protocol (SNMP) and NetFlow~\cite{netflow} are widely used in network management. Passive measurement imposes low or even zero overheads to the network, however, it requires full access to the network devices such as routers and switches. Besides, full access to these devices raises privacy and security issues. As a result, these limitations impede the usage of passive measurements in practice.

The flexibility of SDN yields both opportunities and challenges to monitor the network. Traditional network monitoring techniques such as NetFlow~\cite{netflow} and sFlow~\cite{sflow} support various kinds of measurement tasks, but the measurement and deployment cost are typically high. For example, the deployment of NetFlow consists of setting up collector, analyzer and other services. In contrast, monitoring flow statistics in SDN is relatively light-weight and easy to implement: the central controller maintains the global view of the network, and is able to poll flow statistics from any switch at any time. Furthermore, the boundary between active and passive measurement in SDN is blurred. The controller proactively polls flow statistics and learns active flows by passively receiving notifications from the switches (ofp\_packet\_in and ofp\_flow\_removed message). The challenge is that all the monitoring traffic has to be forwarded to the controller which is likely to result in a bandwidth bottleneck. The situation becomes worse for in-band SDN deployment when monitoring and routing traffic are sharing bandwidth along the same link.

The existing pull-based measurement approaches such as OpenTM~\cite{opentm} utilize many switch selection heuristics to gather the flow statistics. It generates a single query for each source-destination pair to obtain the traffic matrix. If the number of active flows is large, the extra communication cost for each flow cannot be neglected. In order to reduce the monitoring overheads in SDN, FlowSense~\cite{flowsense} is proposed to infer the network utilization by passively capturing and analyzing the flow arrival and expiration messages. However, FlowSense calculates the link utilization only at discrete points in time after the flow expires. This limitation cannot meet the real-time requirement, neither can the accuracy of the results be guaranteed. We argue that the existing approaches are \emph{sub-optimal} as they lack global optimization to choose the polling switches. On the other hand, how to reduce the network consumption for measurement traffic is not well studied by far.

To address the aforementioned issues, we propose \emph{FlowCover}, a low-cost high-accuracy scheme that collects the flow statistics across the network in a timely fashion. Our approach significantly reduces the communication cost of monitoring by aggregating the polling requests and replies. We leverage the global view of SDN to optimize the monitoring strategies. The polling scheme is dynamically changed with real-time traffic across the network. To the best of our knowledge, this is the first work to global optimize the SDN monitoring problem formally.

The primary contributions of our approach are listed below:
\begin{itemize}
\item
  We provide a general framework to facilitate various monitoring tasks such as link utilization, traffic matrix estimation, anomaly detection, etc.
\item
  We introduce a globally optimized flow statistics collection scheme. Our approaches select target switches by the view of all active flows instead of on a per-flow basis.
\item
  Extensive experimental results show that FlowCover reduces roughly $50\%$ monitoring overheads without loss of accuracy in most of the time.
\end{itemize}

The rest of this paper is structured as follows. Section~\ref{sec_motivation} illustrates the motivation of FlowCover by an example. Section~\ref{sec_design} presents the architecture of FlowCover and formulates the problem. Section~\ref{sec_evaluation} elaborates the performance of FlowCover by simulation results. Finally, Section~\ref{sec_relatedwork} summarizes related work and Section~\ref{sec_conclusion} concludes the paper.

\begin{figure}[!t]
  \centering
  \includegraphics[width=2.6in]{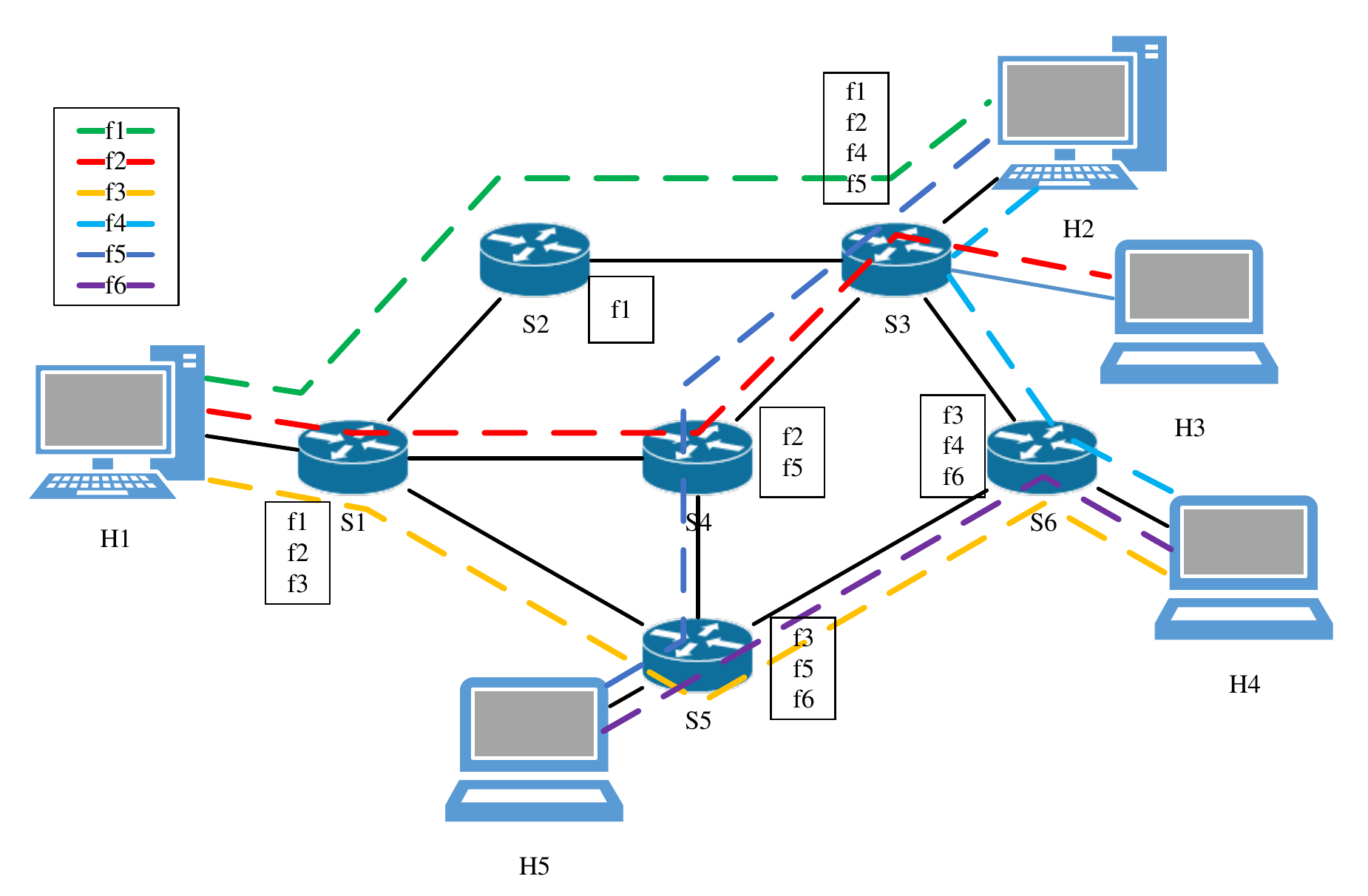}
  \caption{Motivation Example. There are six switches and five hosts in the network. Five active flows are plotted in different colors: $f1: H1-H2; f2: H1-H3; f3: H1-H4; f4: H2-H4; f5: H2-H5; f6: H4-H5$. Each of the switches only holds the partial view of all the flows. The partial view of each switch is given in the rectangles.}
  \label{fig_motivation}
\end{figure}

\section{Motivation} \label{sec_motivation}
OpenFlow~\cite{openflow} is an implementation of SDN. Currently, OpenFlow-based SDN is widely used in both industry and academia. OpenFlow is the de facto standard communication interface between the control plane and data plane. It is an application-layer protocol which contains Ethernet header, IP header and TCP header. According to the OpenFlow specification 1.0~\cite{openflowspec10}, the message body of an individual flow statistics request and reply message has a minimum length of 56 bytes and 108 bytes respectively (at least 1 flow). Therefore, the minimum length of flow statistics request and reply message on wire are 122 bytes and 174 bytes respectively.

Note that the request and reply message are of almost the same length, hence it is promising to design polling schemes to reduce the monitoring overheads, especially in the scenarios with high polling frequency. The key insight is that we \emph{aggregate} the request and reply messages by optimizing the selection of polling switches. The strategy is to intelligently poll a small number of switches which cover a large ratio of flows to minimize the monitoring overheads. For simplicity but without loss of generality, we consider out-of-band deployment of the control network in this paper. An example is shown in Figure~\ref{fig_motivation} to illustrate the problem. 

\begin{figure}[!t]
  \centering
  \includegraphics[width=2.6in]{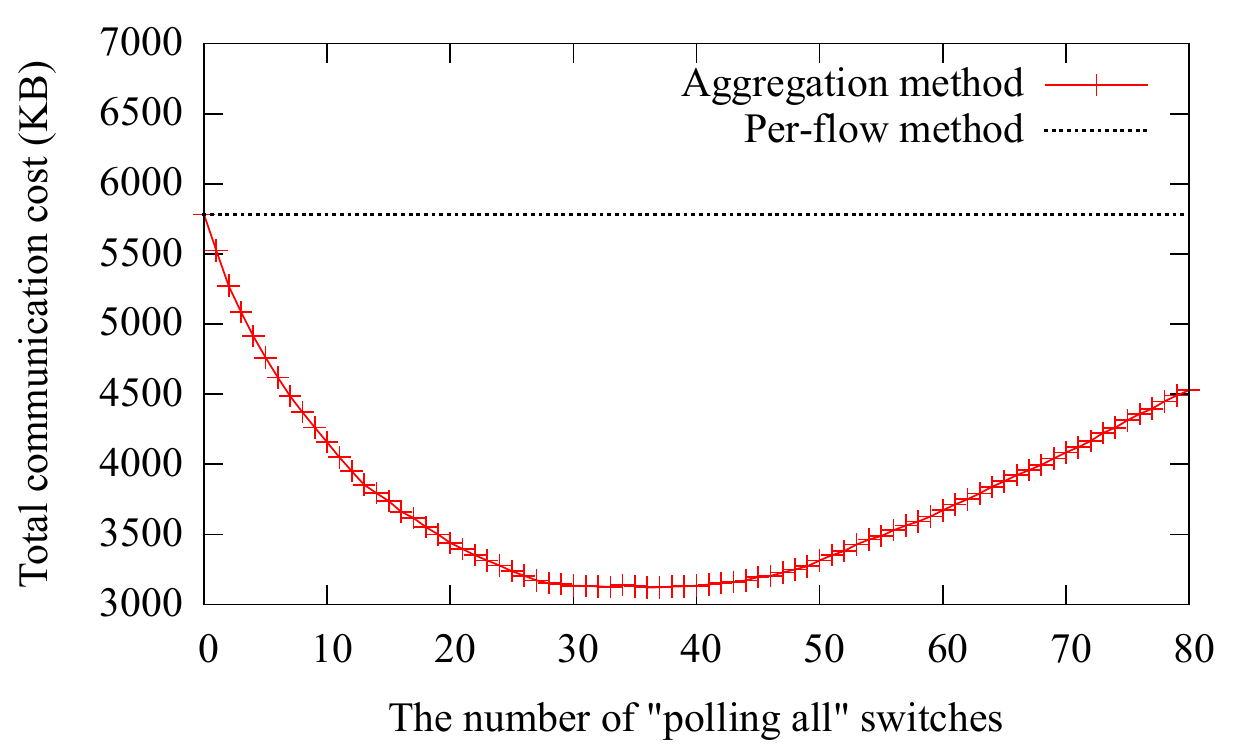}
  \caption{The number of ``polling all'' switches vs. total communication cost in a random graph with $100$ switches and $20000$ active flows in the network.}
  \label{fig_setcover}
\end{figure}

A naive approach to obtain the whole flow statistics is to query one of the switches along the path for each flow and merge the results. However, according to the aforementioned analysis of the length of flow statistics request and reply messages, this strategy imposes too many overheads to the network as it collects statistics on a per-flow basis: repeated request and reply message headers. In order to poll all flow statistics with minimum communication cost, we can design a \emph{globally optimized} strategy to reduce the request messages and aggregate the reply messages.

OpenFlow specification~\cite{openflowspec10} defines a match structure to identify one flow entry or a group of flow entries. A match structure consists of many fields to match against flows such as input switch port, Ethernet source/destination address and source/destination IP. However, it is impractical to select an arbitrary number of flows with ``segmented'' fields due to the limited expression of a single match structure. For instance, consider the four flows passing $S3$, assume the source and destination of these flows are: $f1:(H1,H2)$; $f2:(H1,H3)$; $f4:(H2,H4)$; $f5:(H5, H2)$. Notice that $H1$, $H2$, $H3$, $H4$ belong to different subnets, it is impossible to construct a single match structure to match both $f2$ and $f4$ at the same time. As a result, the polling method is either polling a single flow entry with an exact match structure or polling all flow entries from the switch. In this example, the optimal solution is querying $S3$ and $S6$, with communication cost of $C_{opt}=122+472+122+376=1092$ bytes. Compared with the cost of the naive approach $C_{\text{per-flow}}=(122+174)*6=1776$ bytes, we save about $38.5\%$ of the communication cost. We have much more performance gain in practice as the number of flows and the network scale are much larger than this simple example. In high-accuracy monitoring systems that require high polling frequency, such optimization is of great importance to reduce the monitoring overheads.

Actually, polling flow statistics from one switch as much as possible is a sort of aggregation technique to save the communication cost. However, if this ``polling all'' strategy is employed excessively, it brings extra overheads due to repeated gathering the same flow statistics from different switches. To further explore the problem, we use a simple greedy algorithm which chooses the switches that cover the most number of uncovered flows to collect all the flow statistics. Figure~\ref{fig_setcover} illustrates the trend of total communication cost as the number of ``polling all'' switches varies from $0$ to $80$. The dashed line is the total communication cost of per-flow method for comparison. For aggregation method, there has been a steady fall before the number of ``polling all'' switches reaches $30$. After reaching the bottom, the total communication cost rises gradually until all the active flows have been covered. In summary, our target is to design an efficient model to generate a cost-effective polling scheme that reaches the lowest point of the total communication cost.

\section{System Design and Problem Formulation} \label{sec_design}
In this section, we first give an overview of FlowCover and describe its architecture. The problem formulation and its corresponding heuristics are presented thereafter. A practical algorithm to handle the flow changes is proposed as well.

\subsection{Architecture} \label{sec_architecture}
Basically, the monitoring task in SDN is accomplished by the controller which is connected to all the switches via a secure channel. The secure channel is usually a TCP connection between the controller and the switch. The controller collects the real-time flow statistics from the corresponding switches, and merges the raw data to provide interfaces for upper-layer applications.

\begin{figure}[!t]
  \centering
  \includegraphics[width=2.6in]{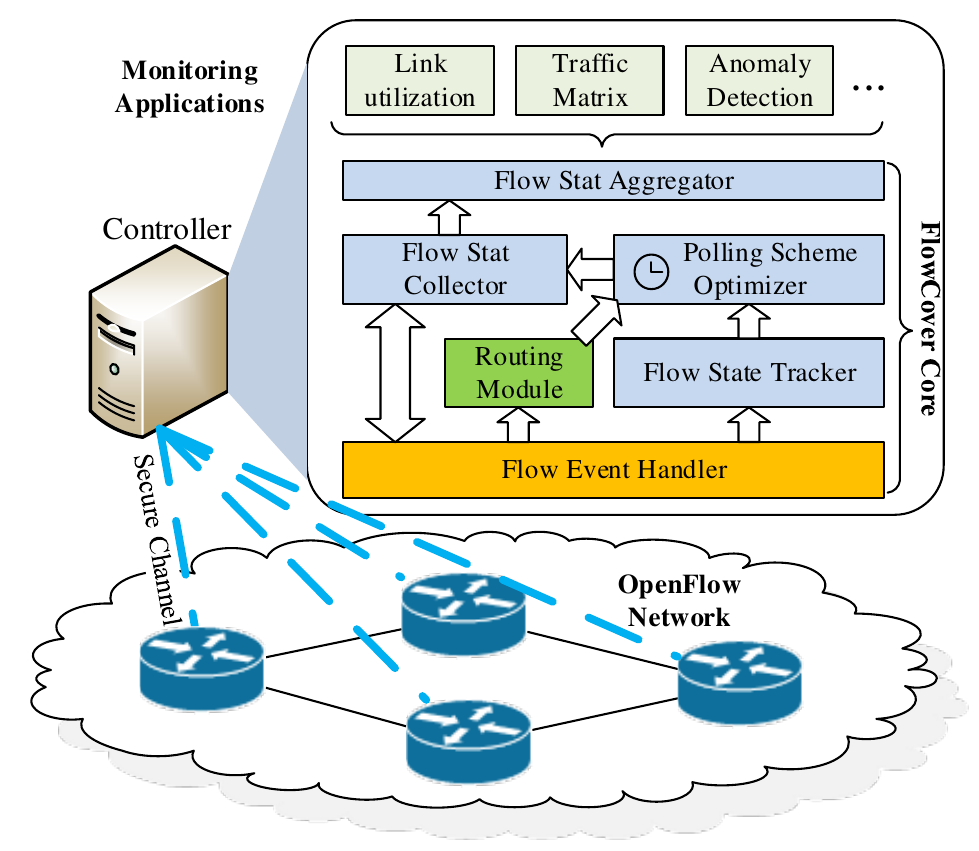}
  \caption{FlowCover architecture}
  \label{fig_architecture}
\end{figure}

We elaborate the architecture of FlowCover in Figure~\ref{fig_architecture}. In general, there are three layers in FlowCover: \emph{OpenFlow Network Layer}, \emph{FlowCover Core Layer} and \emph{Monitoring Applications Layer}. The OpenFlow Network Layer consists of underlying low-level network devices and keeps connections between the controller and the switches. The FlowCover Core Layer is the heart of the monitoring framework. The flow event handler receives the flow arrive/expire messages from switches and forwards them to the routing module and flow state tracker. While the routing module calculates the routing path in terms of the policy defined by the administrator, the flow state tracker maintains the active flows in the network in real-time. The routing module and the flow state tracker report the active flow sets and their corresponding routing paths to the polling scheme optimizer respectively. Based on the above information, the polling scheme optimizer computes a cost-effective polling scheme and forwards it to the flow stat collector. The flow stat collector takes the responsibility to poll the flow statistics from the switches and handle the reply. Finally, the flow stat aggregator gathers the raw flow statistics and provides interfaces for the upper monitoring applications. The Monitoring Applications Layer is a collection of different monitoring tasks such as link utilization, traffic matrix estimation and anomaly detection, etc.

\subsection{Problem Formulation} \label{sec_formulation}
As mentioned in Section~\ref{sec_motivation}, we can poll flow statistics from a switch by two strategies: (1) exact match of one flow; (2) wildcarding all fields to collect all flows. The benefits of the latter strategy is that we reduce the number of request messages and repeated reply headers. On the other hand, excessively usage of the second strategy imposes extra communication cost as there are overlap flow statistics. Therefore, the problem can be formulated as an optimization problem whose objective is to minimize the communication cost.

The target network is an undirected graph $G=(V,E)$, where $V=\{v_1,v_2,\ldots,v_n\}$ is the set of switches and $E$ represents the set of links. Therefore, $n=|V|$ is the number of switches in the network. There are $m$ active flows in the network $F=\{f_1, f_2, \ldots f_m\}$ (called the universe), where each element $f_i,i=1,2,\ldots, m$ corresponds to a sequence of switches $P_i$ that represents the flow routing path with length $l$: $P_i=(v_{j_1},v_{j_2}, \ldots, v_{j_l}), j_q \in [1, n],q \in [1,l]$. Let $l_{req}$ denotes the length of the flow statistics request message, $l_{rh}$ denotes the length of flow statistics reply message header, $l_{sf}$ denotes the length of reply message body of a single flow entry. For a flow statistics reply message with $n$ entries, the whole reply message length $l_{reply}(n)$ is a linear function of $n$\footnote{According to the OpenFlow specification~\cite{openflowspec10}, $l_{req}=122$ bytes, $l_{replyheader}=78$ bytes, $l_{singleflowentry}=96$ bytes.}:
\begin{equation} \label{eq_replymsg}
  l_{reply}(n) = l_{replyheader}+n*l_{singleflowentry}
\end{equation}

Given the network graph $G$ and the active flow set $F$, generate a collection of sets $S1=\{s_1,s_2, \ldots, s_n\}$ that each element $s_i$ is a set of the flows that pass $v_i$. For each $f_i \in F$, generate a single set $s_k=\{f_i\},(k=n+1,n+2,\ldots,n+|F|)$ and add $s_k$ to $S2$. Map the set $s_k$ to $f_i$ and add the mapping to the map $M$, namely $M(s_k)=f_i$. According to (\ref{eq_replymsg}), for each set $s_i, (i=1,2,\ldots,n)$ in $S1$, assign the weight $w_i=l_{req}+l_{reply}(|s_i|)$ to it; for each set $s_i,(i=n+1,n+2,\ldots,n+|F|)$ in $S2$, assign the weight $w_i=l_{req}+l_{reply}(1)$ to it. The algorithm for constructing $S$ from given $G$ and $F$ is shown in Algorithm~\ref{algo_constructsetcover}. The complexity of the transformation is $O(m+n)$. Let $S=S1 \cup S2$, the integer linear programming (ILP) formulation of the problem is:

\begin{equation} \label{eq_problem}
  \begin{split}
    \min & \sum\limits_{s \in S}w_sx_s \\
    \text{subject to:} & \sum\limits_{s: f \in s}x_s \ge 1, \forall f \in F\\
    & x_s \in \{0,1\}, \forall s \in S
  \end{split}
\end{equation}

\begin{algorithm}[!t] \footnotesize
  \caption{Construct Weighted Sets}
  \begin{algorithmic}[1]
    \Function{ConstructWeightedSets}{$G=(V,E), F$}
    \State{$S \gets \{(v_1:\{\}),(v_2:\{\}),\ldots,(v_n:\{\})\}$}
    \State{$W \gets []$} \Comment{$W$ is the weight list for $S$}
    \ForAll{$f \in F$}
    \ForAll{$v \in P_f$}
    \State{$S[v] \gets S[v].append(f)$}
    \EndFor
    \State{$S \gets S \cup \{f\}$} \Comment{Add single flow polling set}
    \EndFor
    \ForAll{$s \in S$}
    \State{$W[s]=l_{req}+l_{reply}(|s|)$}
    \EndFor
    \State{\textbf{return} $S,W$}
    \EndFunction
  \end{algorithmic}
  \label{algo_constructsetcover}
\end{algorithm}

This formulation is the weighted set cover problem that is known to be NP-hard~\cite{approximationbook}. Now we justify how to obtain a polling scheme from the solution of (\ref{eq_problem}). Let $t=n+|F|$, for any solution $X=(x_1,x_2,\ldots, x_t)$, the corresponding polling scheme is:
\begin{itemize}
\item
  $x_{s_i}=1, 1 \le s_i \le n$: Poll all flows from the switch $v_i$.
\item
  $x_{s_i}=1, n+1 \le s_i \le t$: Poll one flow (and it is the only flow in the set) $f \in s_i$ from one of the switch $v \in P_f$\footnote{The strategies to choose the switch have been studied in OpenTM~\cite{opentm}.}.
\end{itemize}

\subsection{Greedy Algorithm}
Since the optimization problem (\ref{eq_problem}) is NP-hard, we propose a heuristic algorithm to solve it efficiently. Our greedy strategy is: choose the most cost-effective switches and remove the covered flows, until all flows are covered. The greedy algorithm is shown in Algorithm~\ref{algo_greedy}. The main loop iterates for $O(n)$ time, where $n=|F|$. The most cost-effective set $s$ can be found in $O(\log m)$ time by a priority queue, where $m=|S|$. So the computational complexity of the algorithm is $O(n \log m)$. The algorithm is proved to be a $H_n=\sum\limits_{i=1}^{n}\frac{1}{i} = O(\log n) $ approximation algorithm~\cite{approximationbook} where $n=\max_{s \in S}|s|$.

\begin{algorithm}[!t] \footnotesize
  \caption{Greedy Select Polling Switches}
  \begin{algorithmic}[1]
    \Function{WeightedSetCover}{$S,W$}
    \State{$C \gets \emptyset; P \gets []$} \Comment{$P$ is the picked set list}
    \While{$C \neq U$}
    \State{Find a set $s \in S$ such that $\frac{W[s]}{|s-C|}$ is minimum}
    \State{$P.append(s)$}
    \State{$C \gets C \cup s$}
    \EndWhile
    \State{\textbf{return} $P$}
    \EndFunction
  \end{algorithmic}
  \label{algo_greedy}
\end{algorithm}

\subsection{Handling Flow Changes}
The active flows in the network change from time to time. From Section~\ref{sec_design}, we know that FlowCover detects the flow changes by the flow state tracker. Intuitively, the polling scheme optimizer has to re-calculate the polling scheme upon receiving flow arrive/expire messages. However, we argue that this is not necessarily true in practice. So we propose a new heuristic to handle flow changes:
\begin{itemize}
\item
  When a new flow arrives: if it has been covered by the current polling scheme, no further actions needed; if not, just add one single flow polling rule to the polling scheme.
\item
  When a flow expires: if this flow is collected by single flow polling, remove it from the polling scheme; if not, no actions.
\end{itemize}

Obviously, the heuristic cannot keep the polling scheme always cost-effective, but it prevents the polling scheme from changing too frequently to impose extra overheads on the controller. To keep the polling scheme up to date, we re-calculate the polling scheme periodically.

\section{Evaluation} \label{sec_evaluation}
In order to evaluate the performance of FlowCover, we build a simulator written in Python to test it from different aspects such as the reduced communication cost, overheads, accuracy and the performance of handling flow changes. Experiments are conducted on a computer with Intel i5-650 3.20 GHz (4 cores) processor and 4G RAM.

\subsection{Communication Cost}
\begin{figure*}[!t]
  \begin{minipage}[b]{0.32\linewidth}
    \centering
    \includegraphics[width=\textwidth]{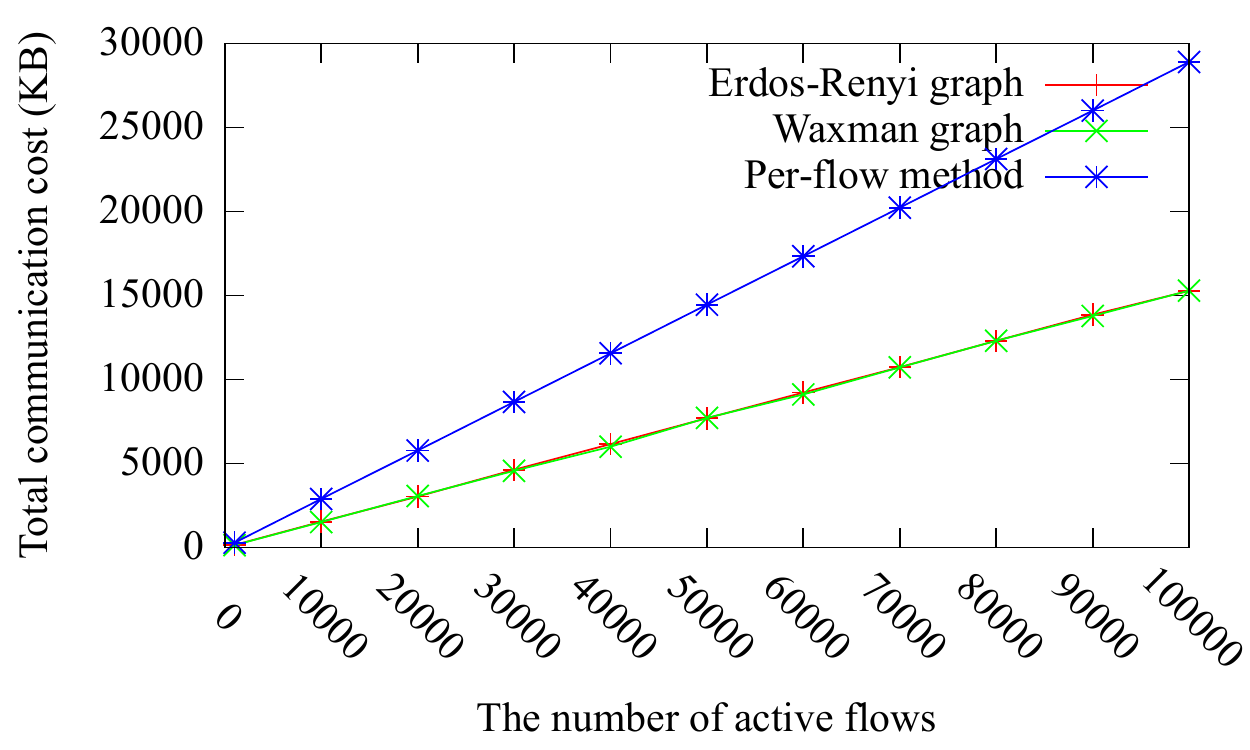}
    \caption{Total communication cost in different graph models.}
    \label{fig_cost}
  \end{minipage}
  \hspace{0.01\linewidth}
  \begin{minipage}[b]{0.32\linewidth}
    \centering
    \includegraphics[width=\textwidth]{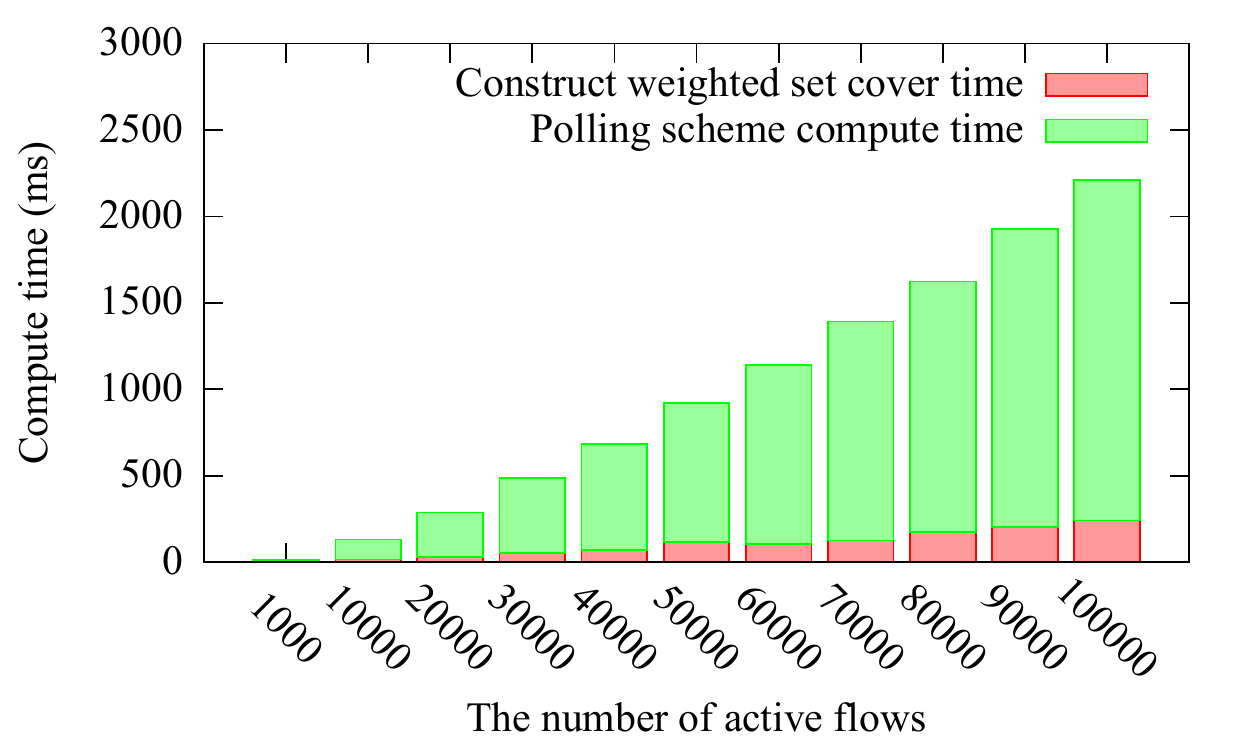}
    \caption{The weighted set cover construction time and solution computing time.}
    \label{fig_constructsetcover}
  \end{minipage}
  \hspace{0.01\linewidth}
  \begin{minipage}[b]{0.32\linewidth}
    \centering
    \includegraphics[width=\textwidth]{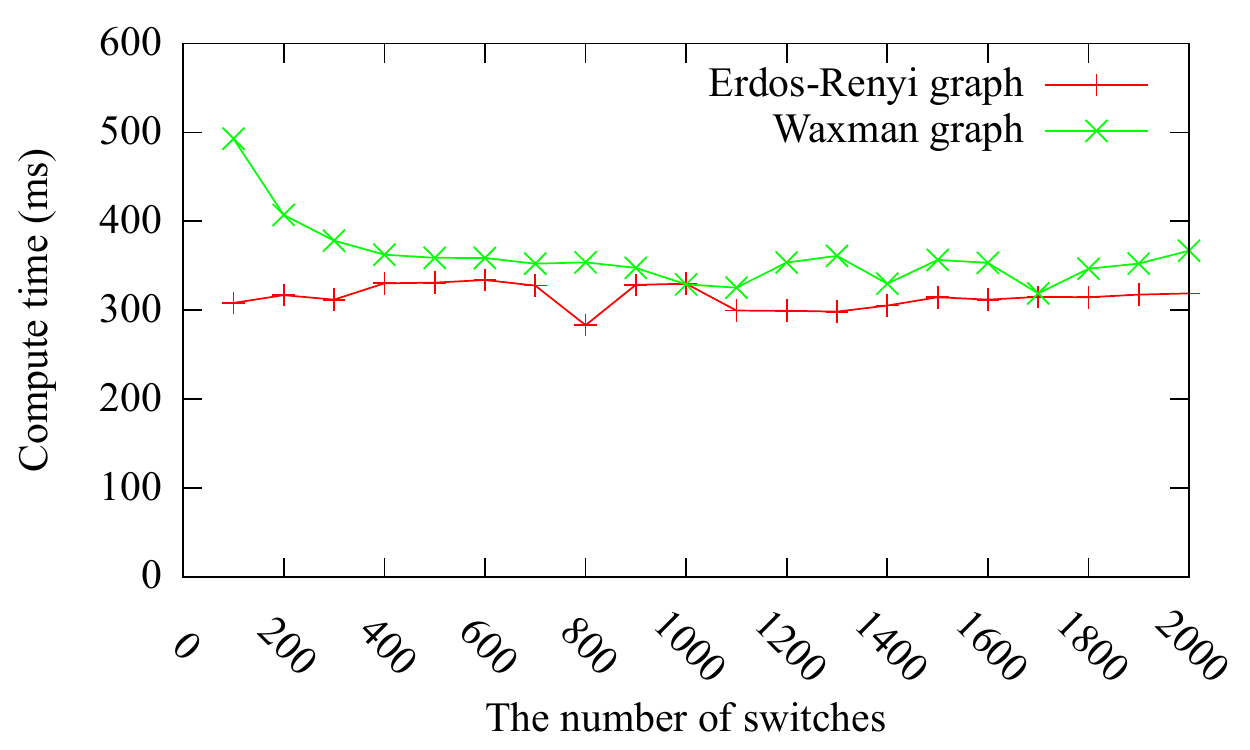}
    \caption{The total computing time vs. the number of switches in the network with $20000$ active flows.}
    \label{fig_caltimevsswitch}
  \end{minipage}
\end{figure*}

We fist elaborate the reduction of communication cost by FlowCover. We compare it with the basic per-flow polling method proposed in~\cite{opentm}. Figure~\ref{fig_cost} shows the total communication cost in Erd\H{o}s-R\'enyi graph~\cite{erdos} and Waxman graph~\cite{waxman} which are widely used in network research. Both of the graphs consists of $200$ switches. We generate flows and choose the source and destination in a uniformly random manner. The number of active flows varies from $1000$ to $100000$ which is a large number for a middle-sized data center. The total communication cost of per-flow polling method is irrelevant to the network topology, because it always generate polling traffic regardless of the flow forwarding path. As a result, we plot only one curve for reference in Figure~\ref{fig_cost}. Compared with the per-flow polling method, FlowCover reduces the total communication cost of the monitoring traffic significantly in both network topologies with different number of active flows. FlowCover saves up to $47.2\%$ of the total communication cost. The experiments illustrate that FlowCover reduces about half of the total communication cost, regardless of the network topology and the number of active flows.

\begin{figure*}[!t]
  \begin{minipage}[b]{0.32\linewidth}
    \centering
    \includegraphics[width=\textwidth]{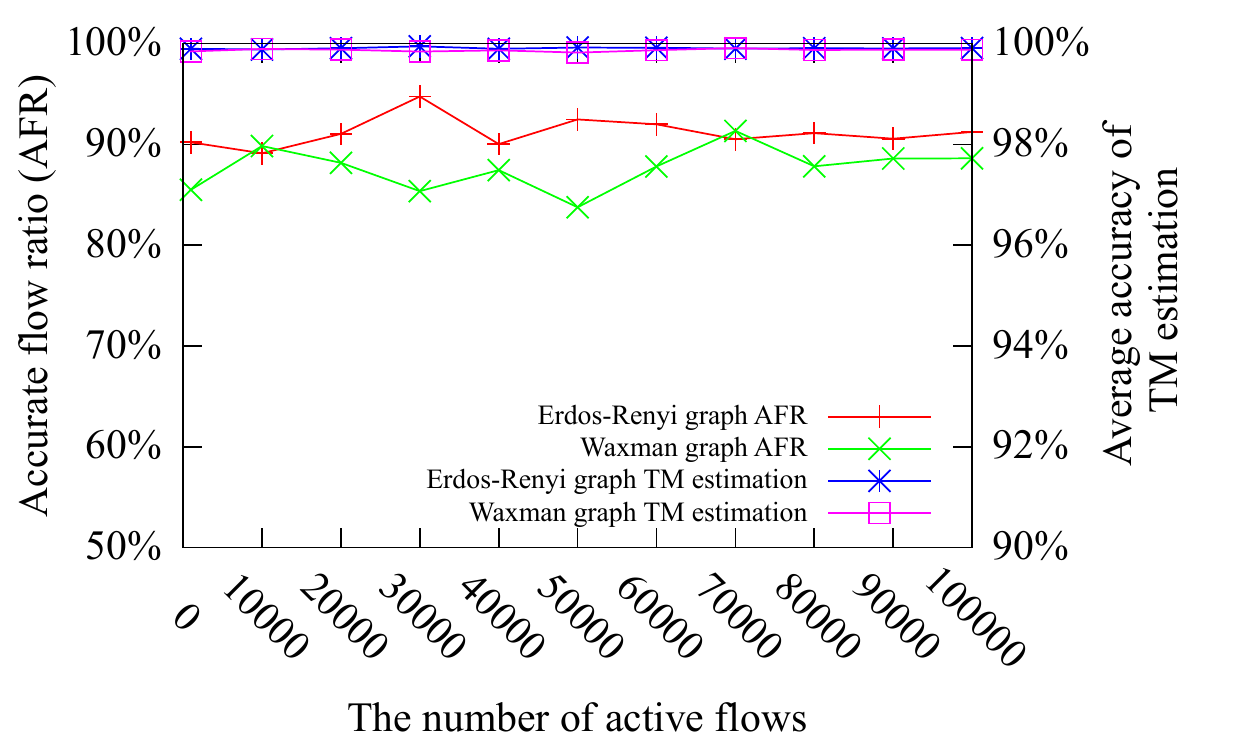}
    \caption{Accurate flow ratio and average accuracy of traffic matrix (TM) estimation vs. the number of active flows in a $200$ switches network. The packet loss rate and loss switch ratio is $1\%$ and $10\%$ respectively.}
    \label{fig_accuracyvsflow}
  \end{minipage}
  \hspace{0.01\linewidth}
  \begin{minipage}[b]{0.32\linewidth}
    \centering
    \includegraphics[width=\textwidth]{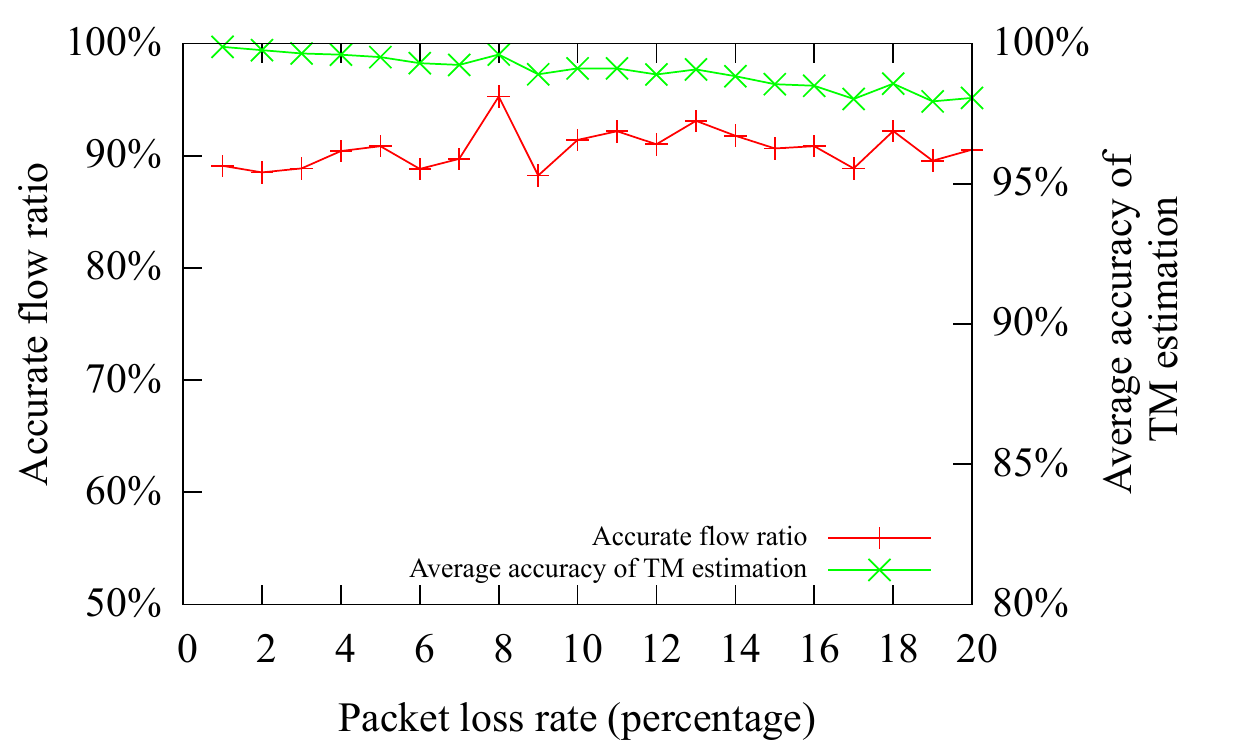}
    \caption{Accurate flow ratio and average accuracy of traffic matrix estimation vs. packet loss rate in a $200$ switches Erd\H{o}s-R\'enyi network. The loss switch ratio is $10\%$.}
    \label{fig_accuracyvspacketloss}
  \end{minipage}
  \hspace{0.01\linewidth}
  \begin{minipage}[b]{0.32\linewidth}
    \centering
    \includegraphics[width=\textwidth]{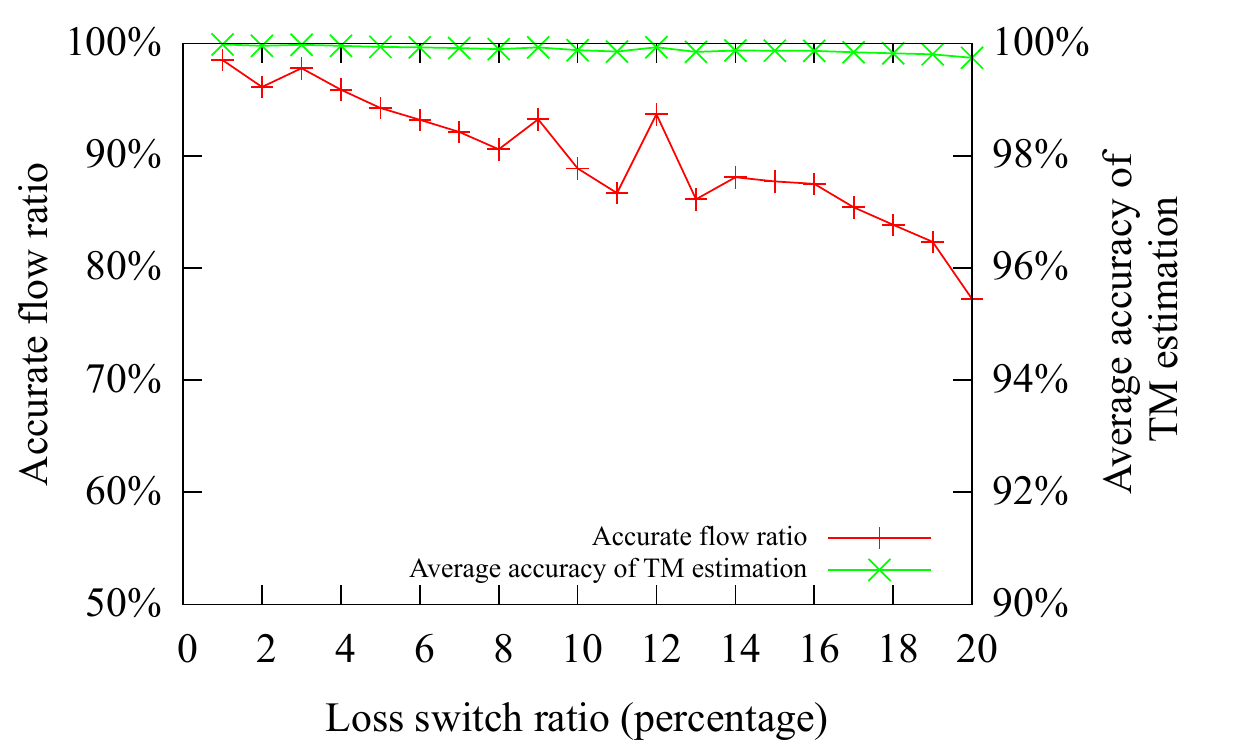}
    \caption{Accurate flow ratio and average accuracy of traffic matrix estimation vs. loss switch ratio in a $200$ switches Erd\H{o}s-R\'enyi network. The packet loss rate is $1\%$.}
    \label{fig_accuracyvslossswitch}
  \end{minipage}
\end{figure*}
\subsection{Overheads}
We examine the overheads of FlowCover, especially on the construction time of the weighted set cover problem and the polling scheme calculation time. The experiment is conducted in a Erd\H{o}s-R\'enyi network with $200$ switches. The result is shown in Figure~\ref{fig_constructsetcover}. There has been a steady increase in the total computing time over the number of active flows. The problem construction time occupies roughly $10\%$ of the total calculation time. The polling scheme computing time is almost linear in the number of active flows (with fixed number of switches) which conforms to the complexity of the greedy algorithm. Our approach obtains the optimized polling scheme very efficiently in practice: for a network with up to $100000$ active flows, we get the polling scheme in less than $2.5s$. The computation time is much less than most of the real-world polling frequency of monitoring tasks.

The relation between the number of switches and the polling scheme computing time is explored as well. As shown in Figure~\ref{fig_caltimevsswitch}, for $20000$ active flows in a Erd\H{o}s-R\'enyi graph, the polling scheme computing time keeps relatively stable, since the computing time is in logarithm relation with the number of switches.

\subsection{Accuracy}
We evaluate the accuracy of FlowCover by two metrics: accurate flow ratio (AFR) which indicates the percentage of accurate flows; average accuracy of traffic matrix (TM) estimation which represents the error between the measured and real traffic matrix. We emulate packet loss in the simulator by introducing two parameters: packet loss rate and loss switch ratio. The switches in the simulator are divided into two categories: normal switch and loss switch. When a packet passes a loss switch, it is dropped with a probability of packet loss rate. Loss switch ratio is defined as the number of loss switches to the number of all switches. We generate loss switches in a uniformly random manner according to the loss switch ratio.

Figure~\ref{fig_accuracyvsflow} shows the AFR and accuracy of TM estimation in different network topologies. The AFR fluctuates around $90\%$ in both topologies. The accuracy of TM estimation always above $99\%$. Figure~\ref{fig_accuracyvspacketloss} illustrates that the AFR resists to the increasing packet loss rate; the accuracy of TM estimation falls gradually from $99.9\%$ to $98.1\%$. Figure~\ref{fig_accuracyvslossswitch} shows that the AFR falls in proportion to the loss switch ratio. However, the accuracy of TM estimation only decreases slightly from $99.9\%$ to $99.7\%$. These experiments demonstrate that FlowCover saves the communication cost without loss of accuracy.

\subsection{Handling Flow Changes}
\begin{figure}[!t]
  \centering
  \includegraphics[width=2.8in]{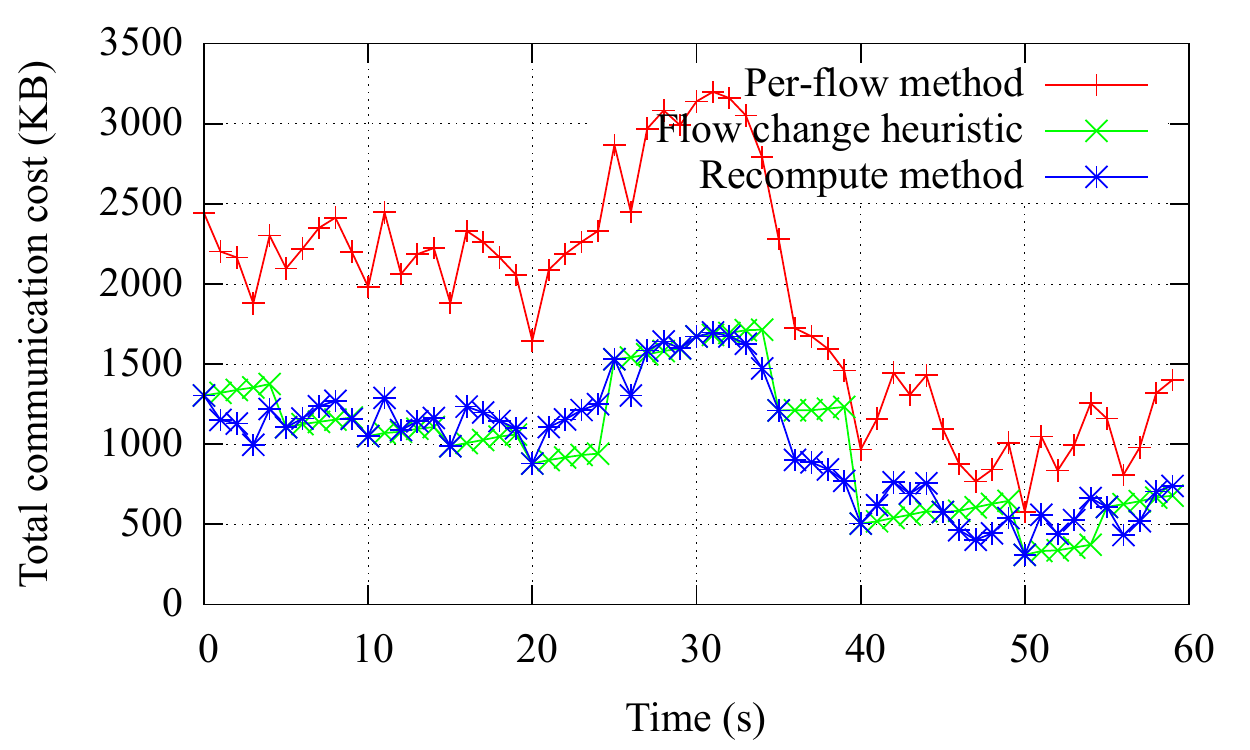}
  \caption{The performance of the flow change heuristic.}
  \label{fig_flowchange}
  \vspace{-0.1in}
\end{figure}

The performance of flow change heuristic is presented in Figure~\ref{fig_flowchange}. The experiment is conducted in a $200$ switches Erd\H{o}s-R\'enyi network with $10000$ active flows initially. We generate random number of flow arrive/expire events from $[0,2000]$ each second. We set the polling frequency and the scheme re-computed interval to $1s$ and $5s$ respectively. The experiment lasts for $1$ minute. The total communication cost of per-flow polling method is plotted as the baseline and the cost is in proportion to the number of active flows. It is easy to see that the flow change heuristic does not increase too much communication cost compared with the always re-compute method. This is because most of the new flows have been covered by the current polling scheme. Sometimes, the performance of the heuristic is even better than the re-compute method. The reason is that the polling scheme is calculated by an approximation algorithm. Increasing a limited number of single polling has little impact on the total communication cost. Therefore, the scheme which is obtained by the flow change heuristics might be better than the re-compute one in a short period of time. In summary, the flow change heuristic further reduces the computing overheads of FlowCover and tackle the flow change issue properly.

\section{Related Works} \label{sec_relatedwork}
As the rapid development of SDN and quality of service routing, network monitoring has been emerged as one of the hot topics recently. OpenTM~\cite{opentm} proposed a traffic matrix estimation system, which obtains flow statistics by different querying strategies. It collects active flow statistics on a one-by-one basis which is not cost-effective. FlowSense~\cite{flowsense} presented a push-based method to measure the network link utilization with zero overhead. However, FlowSense can only obtain the link utilization at discrete points in time with a long delay. It cannot meet the real-time monitoring requirement and extend to other general measurement tasks. PayLess~\cite{payless} proposed an adaptive statistics collection algorithm to trade off between accuracy, timeliness and overheads. Dynamically changing the aggregation granularity~\cite{adaptiveflowcounting, onlineaggregate} is an alternative to detect hierarchical heavy hitter and malicious attacks. However, these approaches use extra wildcard rules to obtain the flow statistics, which waste precious TCAM resources and increase the processing time of packets. Detailed analysis of the trade-off between resource consumption and measurement accuracy has also been studied~\cite{resourceaccuracy}. Besides, reducing extra overhead introduced by SDN is also studied in~\cite{cheetahflow}.

ProgME~\cite{progme} enabled flexible flow counting by defining the concept of flowset that is an arbitrary set of flows for different applications. OpenSketch~\cite{opensketch} provided a three-stage pipeline switch design to accommodate different sketch algorithms. It supports different sketch-based measurement tasks at low implementation cost. DevoFlow~\cite{devoflow} proposed different ways to improve the performance of statistics collection such as sampling, trigger-and-report and approximation counters.


\section{Conclusion and Future Work} \label{sec_conclusion}
In this paper, we propose FlowCover, a low-cost high-accuracy monitoring framework for SDN. We analyze the communication overheads of SDN monitoring and provide a general framework to accommodate various monitoring tasks. We model the polling switches selection as a weighted set cover problem and optimize the communication overheads globally. Heuristics are presented to solve the optimization problem and handle flow changes in practice. Extensive experimental results show that FlowCover reduces roughly $50\%$ of the communication cost in most cases.

Our future work is to extend our scheme to a more general monitoring framework, which takes the switch load, flow forwarding latency and multi-tenant scenario into consideration. In addition, we only consider single controller scenario in this paper, we also plan to make a more scalable, high-performance monitoring framework by designing distributed SDN-based schemes.

\section*{Acknowledgment}
This research is supported by HKUST Research Grants Council (RGC) 613113.

\bibliographystyle{IEEEtran}
\bibliography{references}

\begin{thebibliography}{10}
\providecommand{\url}[1]{#1}
\csname url@samestyle\endcsname
\providecommand{\newblock}{\relax}
\providecommand{\bibinfo}[2]{#2}
\providecommand{\BIBentrySTDinterwordspacing}{\spaceskip=0pt\relax}
\providecommand{\BIBentryALTinterwordstretchfactor}{4}
\providecommand{\BIBentryALTinterwordspacing}{\spaceskip=\fontdimen2\font plus
\BIBentryALTinterwordstretchfactor\fontdimen3\font minus
  \fontdimen4\font\relax}
\providecommand{\BIBforeignlanguage}[2]{{%
\expandafter\ifx\csname l@#1\endcsname\relax
\typeout{** WARNING: IEEEtran.bst: No hyphenation pattern has been}%
\typeout{** loaded for the language `#1'. Using the pattern for}%
\typeout{** the default language instead.}%
\else
\language=\csname l@#1\endcsname
\fi
#2}}
\providecommand{\BIBdecl}{\relax}
\BIBdecl

\bibitem{netflow}
``{NetFlow},''
  \url{http://www.cisco.com/c/en/us/products/ios-nx-os-software/ios-netflow/index.html}.

\bibitem{sflow}
M.~Wang, B.~Li, and Z.~Li, ``{sFlow: Towards resource-efficient and agile
  service federation in service overlay networks},'' in \emph{ICDCS}, 2004.

\bibitem{opentm}
A.~Tootoonchian, M.~Ghobadi, and Y.~Ganjali, ``{OpenTM: traffic matrix
  estimator for OpenFlow networks},'' in \emph{PAM}, 2010.

\bibitem{flowsense}
C.~Yu, C.~Lumezanu, Y.~Zhang, V.~Singh, G.~Jiang, and H.~V. Madhyastha,
  ``{FlowSense: monitoring network utilization with zero measurement cost},''
  in \emph{PAM}, 2013.

\bibitem{openflow}
N.~McKeown, T.~Anderson, H.~Balakrishnan, G.~Parulkar, L.~Peterson, J.~Rexford,
  S.~Shenker, and J.~Turner, ``{OpenFlow}: enabling innovation in campus
  networks,'' \emph{SIGCOMM CCR}, 2008.

\bibitem{openflowspec10}
``Openflow switch specification 1.0.0,''
  \url{https://www.opennetworking.org/images/stories/downloads/sdn-resources/onf-specifications/openflow/openflow-spec-v1.0.0.pdf}.

\bibitem{approximationbook}
V.~V. Vazirani, \emph{Approximation algorithms}.\hskip 1em plus 0.5em minus
  0.4em\relax Springer-Verlag New York, Inc., 2001.

\bibitem{erdos}
B.~Bollob\'{a}s, \emph{Random graphs}.\hskip 1em plus 0.5em minus 0.4em\relax
  Academic Press, 1985.

\bibitem{waxman}
B.~M. Waxman, ``Routing of multipoint connections,'' \emph{Selected Areas in
  Communications, IEEE Journal on}, 1988.

\bibitem{payless}
S.~R. Chowdhury, M.~F. Bari, R.~Ahmed, and R.~Boutaba, ``{PayLess: A Low Cost
  Network Monitoring Framework for Software Defined Networks},'' in
  \emph{NOMS}, 2014.

\bibitem{adaptiveflowcounting}
Y.~Zhang, ``An adaptive flow counting method for anomaly detection in sdn,'' in
  \emph{CoNEXT}, 2013.

\bibitem{onlineaggregate}
L.~Jose, M.~Yu, and J.~Rexford, ``Online measurement of large traffic
  aggregates on commodity switches,'' in \emph{HotICE}, 2011.

\bibitem{resourceaccuracy}
M.~Moshref, M.~Yu, and R.~Govindan, ``{Resource/accuracy tradeoffs in
  software-defined measurement},'' in \emph{HotSDN}, 2013.

\bibitem{cheetahflow}
Z.~Su, T.~Wang, Y.~Xia, and M.~Hamdi, ``{CheetahFlow: Towards Low Latency
  Software-Defined Network},'' in \emph{ICC}, 2014.

\bibitem{progme}
L.~Yuan, C.-N. Chuah, and P.~Mohapatra, ``{ProgME}: towards programmable
  network measurement,'' \emph{IEEE/ACM Trans. Netw.}, 2011.

\bibitem{opensketch}
M.~Yu, L.~Jose, and R.~Miao, ``{Software defined traffic measurement with
  OpenSketch},'' in \emph{NSDI}, 2013.

\bibitem{devoflow}
A.~R. Curtis, J.~C. Mogul, J.~Tourrilhes, P.~Yalagandula, P.~Sharma, and
  S.~Banerjee, ``{DevoFlow: scaling flow management for high-performance
  networks},'' in \emph{SIGCOMM}, 2011.

\end{thebibliography}

\end{document}